\documentclass[aps,prl,reprint,superscriptaddress]{revtex4-2}
\usepackage{amsmath,amssymb}
\usepackage{graphicx}
\usepackage{hyperref}

\begin{document}
\title{From Flow to Form: Emergence of the Cytokinetic Ring via Active Cortical Dynamics}
\author{Sabyasachi Mukherjee}
\author{Anirban Sain}
\email{asain@phy.iitb.ac.in}
\affiliation{Physics Department, Indian Institute of  Technology-Bombay, Powai, Mumbai, 400076, India. }

\begin{abstract} 
Active cortical flows are essential for assembling the equatorial actin ring during cell division 
and for driving cleavage furrow formation.
However, neither the filamentous ring nor realistic 
furrow shapes have been reproduced in theoretical or computational models. Here, using a phase field 
simulation of an active deformable shell, which 
captures coupling of cortical  flows and nematic 
order, we show how a nematic-like actomyosin ring spontaneously emerge at the equator and drive 
sharp invagination. Unexpectedly, we find that counter-rotating flow patterns near the furrow—often attributed to intrinsic chirality of actomyosin filaments—can instead arise from random initial nematic alignment with a small bias, revealing a memory effect in the cortex. Finally, we propose a reduced model of activity-gradient-driven compressive flow on a flat surface and  identify the minimal physical ingredients 
required for surface instability and opposing flows. Together, these results establish a unified 
physical framework for cytokinesis and provide a mechanistic reinterpretation of flow chirality 
in dividing cells.
\end{abstract}
\maketitle
Cell division is fundamental to biological morphogenesis—a mechano-chemical process through which complex physical shapes and patterns emerge from a single cell \cite{growthandform}. During cytokinesis, the last stage of division, invagination occurs at the midcell, due to flows \cite{reymann2016cortical} in the subsurface
layer - the cortex, which is a contractile, gel-like network of actin filaments and myosin motors. Acto-myosin contractility, which is an active process fueled by hydrolysis of ATP molecules, creates stress gradient that drives active flows \cite{RMP}. \\

A hallmark of cytokinesis is the formation of a contractile actin ring that drives invagination at the mid-cell. 
The resulting division furrow resembles the sharp circular junction of two intersecting spheres \cite{oegema}, where  Gaussian curvature diverges due to a discontinuity in surface normals. 
Assuming isotropic contractility of the active components—and ignoring the filamentous (nematic) character of acto-myosin network—existing theoretical models have thus far reproduced a diffuse, ring-like contractile zone at the mid-cell, along with a dumbbell-shaped furrow exhibiting negative Gaussian curvature \cite{vijay,l2-instability}.
 Recent experiments have further revealed strong azimuthal flows near
 the furrow, often attributed to intrinsic cortical chirality \cite{reymann2016cortical, naganathan-chiral, pimpale2020cell}. Here, using phase-field simulations of an active shell model \cite{julia-prl}, we show that, a) such azimuthal flows, can emerge even in the absence of intrinsic chirality but with a small nonzero mean in the initial random nematic orientation, and b) flow
induced nematic alignment can generate contractile ring leading to sharp invagination at the mid-cell.\\ 

Generic mechano-chemical instabilities of active, deformable, 
closed interfaces have been studied before \cite {mietkePNAS, 
l2-instability, turlier,voigt-royalsocA,RuiMa}, where an 
thin film was coupled to reaction-diffusion dynamics 
of a active scalar concentration field (stress regulators). At high Peclet number instabilities led to a diffuse ring-like accumulation of the stress regulators and a dumbbell shaped invagination, at the equator, via instability of the $l=2$ mode of the spherical harmonics \cite{mietkePNAS, 
l2-instability}. In contrast, stress regulators in our phase field model are nematics, mimicking acto-myosin filaments, which generate anisotropic active stresses. Interestingly, linear stability analysis in Ref\cite{RuiMa}, which considered  nematics, found possibility of transient chiral flow, but for extensile activity, instead of contractile.

 We simulate active cortical flows ${\bf u(r)}$ on a deformable shell, representing the cell cortex \cite{julia-prl,julia-pre}. A phase variable 
$\phi ({\bf r})$ separates two isotropic fluids, the cytoplasm in the cell-interior, at $\phi=1$, and the extra-cellular fluid, at $\phi=-1$. A thin layer (shell) of active gel, representing the cell cortex, lies at the interface.  Orientation distribution of acto-myosin filaments in the cortex is modeled by a nematic order parameter field $Q_{ij}({\bf r})=\langle n_in_j -\delta_{ij}/3\rangle$, where ${\bf \hat n(r)}$ is the nematic director field. In addition, the deformable shell is rendered with Helfrich type \cite{julia-prl} elasticity representative of the cell membrane surrounding the cortex.

\begin{figure}
\centering
\includegraphics[width=9cm, height=7cm]{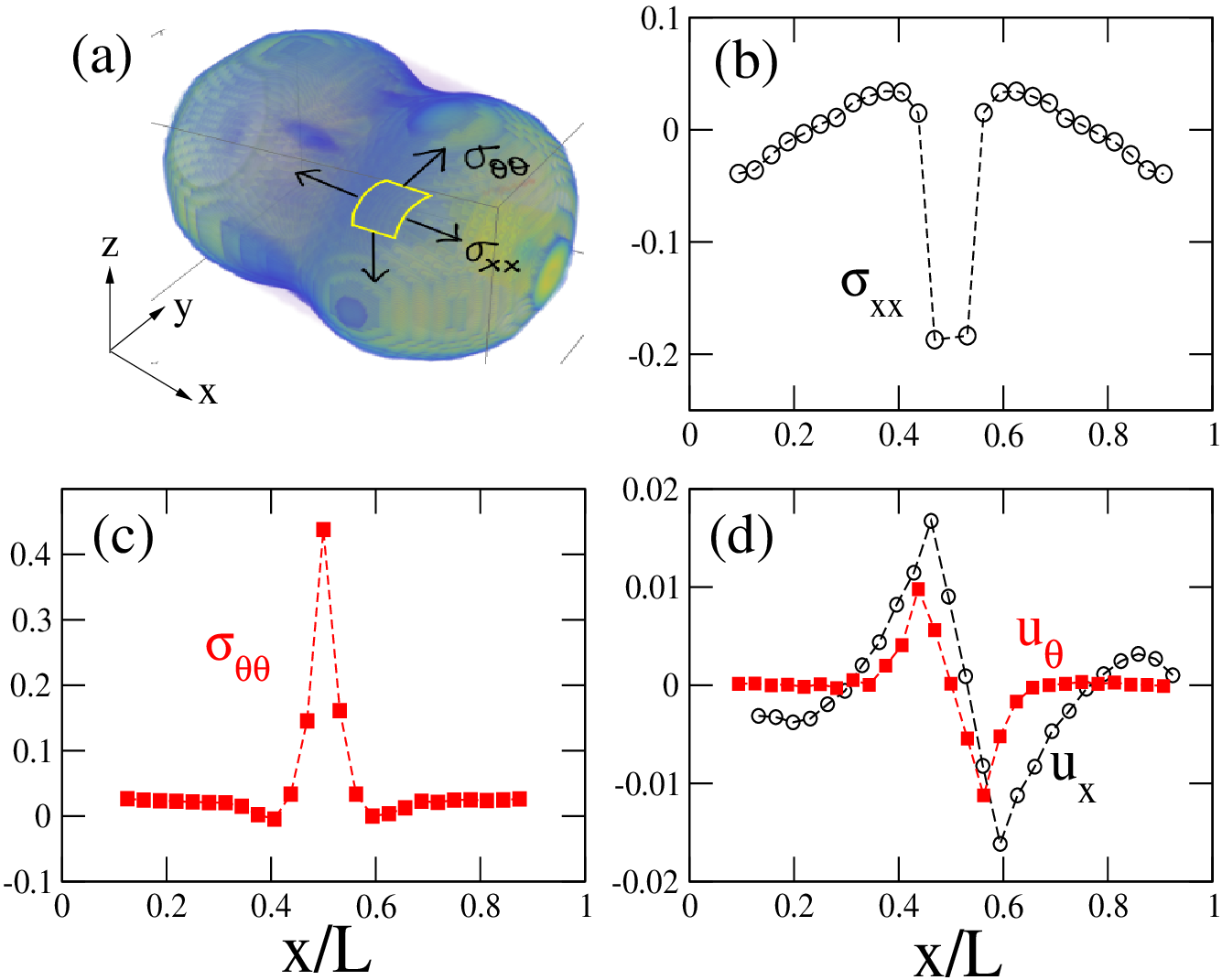}
    \caption{
    Stresses and flows: (a) Shows cell interface, defined by  $|\phi|\leq 0.3$, during invagination and a surface element on it. (b) Shows azimuthally averaged axial stress 
    $\sigma_{xx}(x)$. Its gradient (the local force density) is positive/negative on the left/right lobes respectively, and drives average flow towards the mid-cell.  (c) Sharp rise of $\sigma_{\theta\theta}(x)$ at 
    $x/L=1/2$ implies high contractile ring tension. (d) Shows azimuthally averaged $u_x$ (black, open circles) and $u_{\theta}$ (red, filled squares). Note the significant azimuthal velocity  $u_{\theta}$ near the furrow, indicating counter-rotating flows on the two lobes. 
    }
    \label{fig.stress}
    \end{figure}

The driving force of the cortical flow is the active stress gradient created by the contractile actomyosin gel,  $\sigma^{a}_{ij} =\zeta({\bf r}) Q_{ij}({\bf r},t)$, where the
amplitude $\zeta>0$ and is  assumed to be proportional to the local myosin concentration. Following experimental observation that,  myosin concentration is peaked  at the division furrow \cite{reymann2016cortical}, we impose a static profile $\zeta({\bf r})=\zeta_0 \exp[-(x-\frac{L}{2})^2/2\beta ^2]$, varying only along the division axis $x$ (see Fig.\ref{fig.stress}-a) and peaked at the division plane $x=L/2$. Here $L$ is the length of the simulation box (a cube) and standard deviation $\beta\ll L/2$. However  $\sigma^{a}_{ij}$ still depends on the local nematic alignment $Q_{ij}({\bf r},t)$.

The equations governing the dynamics of the phase field, nematic field and the actomyosin velocity field  are as follows \cite{julia-prl, julia-prx-extraterms},
\begin{eqnarray}
\label{eq.phi1}
&&\frac{\partial \phi}{\partial t} + \mathbf{\nabla} \cdot (\phi \mathbf{u}) = \Gamma_{\phi} \nabla^2 \mu\;.
\\
\label{eq.Q} 
&&\left(\frac{\partial}{\partial t} + \mathbf{u \cdot \nabla}\right)\mathbf{Q} = \mathbf{S} + \Gamma_Q {\mathbf H},\\
\label{eq.u}
&&\rho \left(\frac{\partial}{\partial t} + \mathbf{u} \cdot \nabla\right)\mathbf{u} = -\mathbf{\nabla} p + {\bf \nabla \cdot (\sigma} ^p + {\bf \sigma} ^a)
\\
&&\mbox{along with incompressibility:}\; 
\nabla \cdot \mathbf{u} = 0 \;,\nonumber
\end{eqnarray}
 The phase field in Eq.\ref{eq.phi1} follows a Cahn-Hillard type advection diffusion dynamics \cite{cahn-hillard1958} where the chemical potential 
$\mu=\delta F/\delta\mathbf \varphi$ and the free energy  $F=\int d^3r f(\phi,{\bf \nabla}\phi, {\bf Q,\nabla Q})$ has contribution from various sources explained below. 
The orientation field $Q_{ij}$ in Eq.\ref{eq.Q} is 
governed by fluid advection, the flow coupling $S$ and 
the molecular field
${\bf H} = -(\delta F/\delta {\bf Q} - \frac{1}{3} \mathrm{tr}[\delta F/\delta {\bf Q}] {\bf I})$.
Here $\Gamma_{\phi}$ and $\Gamma_Q$ are the respective relaxation constants.

The free energy density $f$ consists of $Q$ dependent bulk nematic energy $f_b$, orientation gradient ($\nabla Q$) dependent $f_{elastic}$, membrane contribution $f_{mem}$ and surface anchoring term $f_{anchor}$.
\begin{equation}\label{eq:bending_energy}
f_b = A_0 \left[\frac{1}{2} (1 - \frac{\eta(\phi)}{3}) \mathrm{tr}Q^2 - \frac{\eta(\phi)}{3} \mathrm{tr}Q^3 + \frac{\eta(\phi)}{4} (\mathrm{tr}Q^2)^2\right]
\end{equation}
This form \cite{julia-prl} exhibits a first order isotropic-nematic transition at $\eta=2.7$, when we choose $\eta(\phi) = \eta_0 - \eta_s(\phi - \bar{\phi})^2$. With $A_0>0$ and $\bar \phi=0$, a nematic phase can be induced at the interface of two isotropic fluid phases, corresponding to the minima $\phi=\pm 1$, which represent the interior and exterior of the cell. The membrane energy is
\begin{equation}\label{eq:membrane_energy}
f_{\text{mem}} = \frac{\kappa}{2} \left(- \phi + \phi^3 - \epsilon^2 \nabla^2 \phi\right)^2 + \frac{k_{\phi}}{2} (\nabla \phi)^2\;,
\end{equation}
which in the sharp interface limit $(\epsilon\rightarrow 0)$  reduces to
the Helfrich free-energy for membranes \cite{campelo-mem, 
sharp-interface}, with both bending and gaussian moduli $\propto \kappa$, and membrane tension  $\propto \sqrt k_{\phi}$. Further, $f_{\text{elastic}} = (L/2)({\bf \nabla Q})^2\), is the Frank free-energy which penalizes orientational gradients. 
An additional term $f_{\text{anchor}} = L_0 {\bf \nabla} \phi \cdot {\bf Q} \cdot {\bf \nabla} \phi$, with $(L_0 > 0)$, ensures that the nematic directors lie parallel to the interface.\\

Eq.\ref{eq.u} is the momentum balance equation. Other than advecting both $\phi$ and $Q$ fields, gradients of the velocity field ${\bf u}$ align the nematics $Q$ through flow coupling term, $
{\bf S} = (\xi {\bf D + \Omega})({\bf Q} + \frac{1}{3}{\bf I}) + ({\bf Q} + \frac{1}{3}{\bf I})(\xi {\bf D - \Omega}) - 2\xi({\bf Q} + \frac{1}{3}{\bf I})\mathrm{tr}({\bf QW})$.
Here ${\bf D}$ and ${\bf \Omega}$ are the symmetric and
the antisymmetric parts, respectively, of the velocity
gradient tensor $W_{ij}=\partial_iu_j$, and $\xi$ is the flow coupling parameter. In addition to myosin dependent
active stress $\sigma^a$, the passive stress ${\bf \sigma}^p$ in Eq.\ref{eq.u} includes the influences of $\phi$
and $Q$ fields on the velocity field. ${\bf \sigma}^p$ consists of three parts,  $\mathbf{\sigma}^p=\mathbf{\sigma}^{vis}+  \mathbf{\sigma}^{elastic} + \mathbf{\sigma}^{cap}$.
The viscous stress is $\mathbf{\sigma}^{vis} = 2\eta \mathbf{D}$, with $\eta$ the fluid viscosity. 
The elastic stress due to the nematic field is given by $\mathbf{\sigma}^{\text{elastic}} =-P_0{\bf I} - \xi \mathbf{H}(\mathbf{Q} + \tfrac{1}{3}\mathbf{I})
- \xi (\mathbf{Q} + \tfrac{1}{3}\mathbf{I})\mathbf{H}
+ 2\xi (\mathbf{Q} + \tfrac{1}{3}\mathbf{I}) \operatorname{tr}(\mathbf{QH})
+ \mathbf{QH} - \mathbf{HQ} - \nabla \mathbf{Q} \frac{\delta f}{\delta \nabla \mathbf{Q}}$, 
where $P_0$ is the bulk pressure. Finally, the capillary stress originating at the interface is $\mathbf{\sigma}^{\text{cap}} = (f - \mu \phi)\mathbf{I} - \mathbf{\nabla} \phi (\delta F / \delta \mathbf{\nabla} \phi).$
\\

\begin{figure*}
\centering
\includegraphics[width=0.9\linewidth]{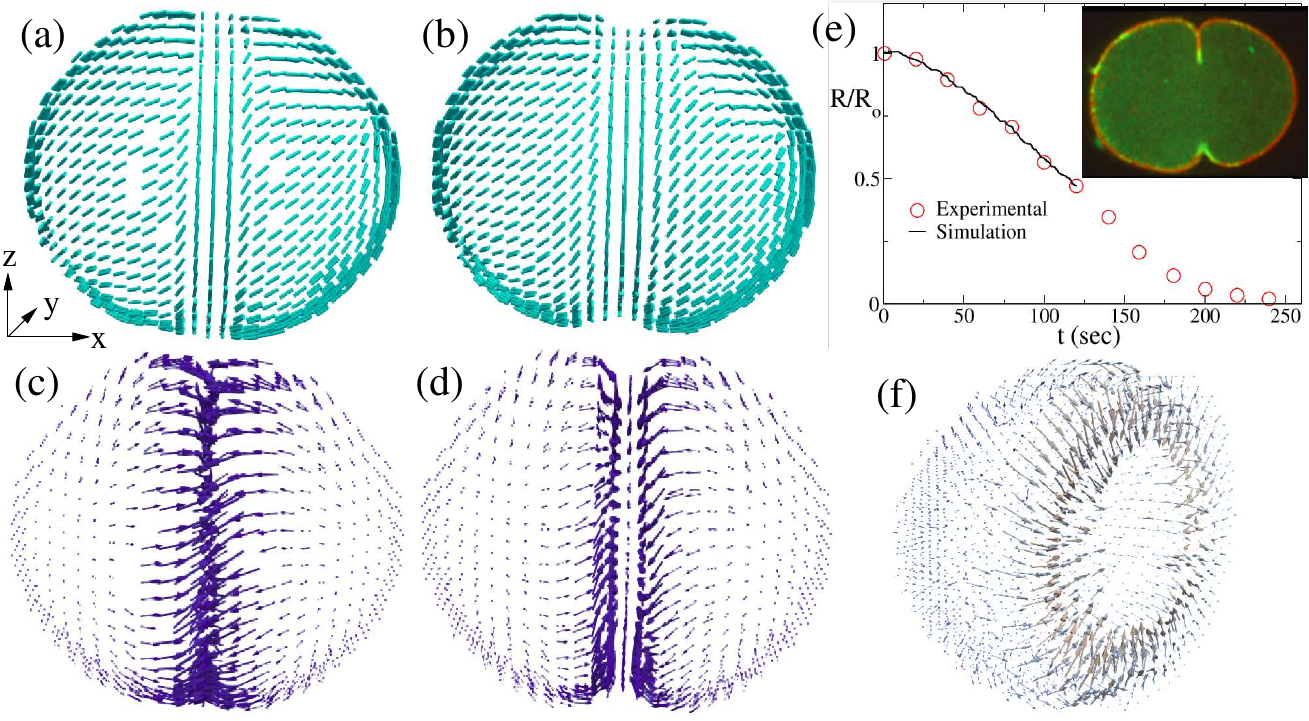}
\caption{Time evolution of nematic field (a$\rightarrow b$) and the corresponding velocity field (c$\rightarrow d$), on active deforming interface 
when started from a sphere of radius $R_0$. Images (a,c) and (b,d) are captured at $5.5\%$ 
and $11\%$ reduction of the furrow radius (closure), respectively. 
A sharp and growing invagination, and strong nematic alignment (the ring) are visible at the equator.
Magnitude of the flow (arrow lengths), in C and D, increases towards the equator (quantified in Fig.\ref{fig.stress}-d). 
The flow also develops an emergent, azimuthal component 
near the furrow.
(E) Fractional change of the furrow radius $R(t)/R_0$ with time (solid line), obtained from our simulation, is compared with 
experimental data ($n=5$) \cite{exptdata5set, sain-julicher} till about $\sim 53\%$ closure (see discussion). 
The inset shows sharp invagination seen in experiments on C. elegans embryo at the single cell stage \cite{oegema} (with permission) during its first division. (F) One lobe of the cell (in our simulation) is shown from the side to highlight the inward flow at the invagination.
} 
\label{fig.evolve}
\end{figure*}

We simulate equations (1)-(3) using psuedo spectral method on a
cubic grid of size $N^3$ ($N=32$ and $64$). The
nonlinear terms are integrated using an Adams-Bashforth scheme \cite{Adams} while the dissipative terms are integrated exactly, using integrating factor.  
The simulation highlights the importance of the nonlinear terms in Eq.1-3; we checked that the linearized version of these PDEs' which 
are often used in analytic calculations \cite{salbreux, mainak-PRL} do not reproduce the invagination phenomenon. The time evolution preserves phase volume ($\dot\phi({\bf k})=0$), consistent with the observation that the cell volume remains constant during cytokinesis (see SI). \\

Our simulation focuses on the first half of invagination, i.e., 
up to $\sim 50\%$ reduction in the furrow radius. Phenomenologically, adhesion bonds stitch the invaginated surfaces of the two daughter cells, keeping the invagination unchanged at the outer surface (see Fig.\ref{fig.evolve}-e), while the 
actin ring leads at the inner edge of the separating surface. 
Fig.\ref{fig.evolve} shows evolution of the nematic orientation field $Q \;(a\rightarrow b)$  and the velocity field ${\bf u}\; (c\rightarrow d)$ on our evolving "cell" surface, defined by a thin shell with phase $|\phi| \leq 0.3$, where $|\vec\nabla\phi|$ is maximum. Also see movies "Mov1" and "Mov2" for the nematic and velocity evolutions, respectively. 
The initial condition (at $t=0$) for $\phi$ was a sphere (with $\phi=+1$ inside and $\phi=-1$ outside). 
The initial nematic field $Q_{ij}$ was random with a nonzero mean inclination with the equator while the initial velocity was set to ${\bf u}=0$ everywhere. 
Results from other initial nematic orientations, including zero mean random orientations, are described in SI. Variety of flow structures, including azimuthally symmetric flows, also emerged. 
As mentioned before, we impose a static axial activity gradient $\zeta(x)$, peaked at the equator, which builds up gradient in axial stress $\sigma_{xx}$ (see Fig.\ref{fig.stress}-a,b) and generates axial flow $u_x$ towards the equator (see Fig.\ref{fig.evolve}-c,d and Fig.\ref{fig.stress}-d).
The stresses and the velocities in Fig.\ref{fig.stress} -b,c,d, shown as functions of $x$,  are azimuthally averaged.\\

Interestingly, strong nematic alignment, parallel to the equator-a ring emerges at the mid-cell (see Fig.\ref{fig.evolve}-a,b). The ring, in turn produces sharply localized, contractile stress $\sigma_{\theta\theta}$ (see Fig.\ref{fig.stress}-a,c) generating a sharp surface invagination. 
The corresponding flow, near the equator, exhibits significant azimuthal component $u_{\theta}$ (see Fig.\ref{fig.stress}-d). 
Away from the ring, the nematic orientation field also exhibits inclination with respect to the equator and this turns out to be important for generating chiral flows, as our analysis shows later. Emergence of the azimuthal component $u_{\theta}$, even in the absence of intrinsic chirality, is the highlight of this work. 
While Fig-1 exhibits no nematic defect near the ring, two aster like +1 defects do form at the cell poles. For other possible flow profiles and nematic arrangements see SI. 

Further, the contractile stress 
$\sigma_{\theta\theta}$ at the equator also gives rise to radially inward velocity $u_r$ at the invaginating plane (see Fig.\ref {fig.evolve}-e and the SI). While both $u_x, u_{\theta}$ tend to zero at the equator, $u_r$ is responsible for advancing the furrow constriction inward. 
Note that the velocity field on the interface has 
both tangential and normal components. The normal component is responsible for transverse movement of the interface as the velocity field advects the phase field. 
In this model, the nematic activity, localized at the $\phi\simeq 0$ interface (the cortex), drives cortical flow, which in turn drives flows (via shearing) in the interior (cytoplasm) and exterior region ($\phi=\pm 1$). Interestingly, we obtain toroidal vortices in the cytoplasm similar to that in Ref.\cite{l2-instability} (see SI). 
The flow in the extracellular region ($\phi=-1$) does not have any physical significance and can be suppressed using a $\phi$ dependent additional viscosity term in Eq.\ref{eq.u}. However, this does not alter the cortical flow pattern or the equatorial invagination.\\

\begin{figure}
\centering
\includegraphics[width=8cm, height=5cm]{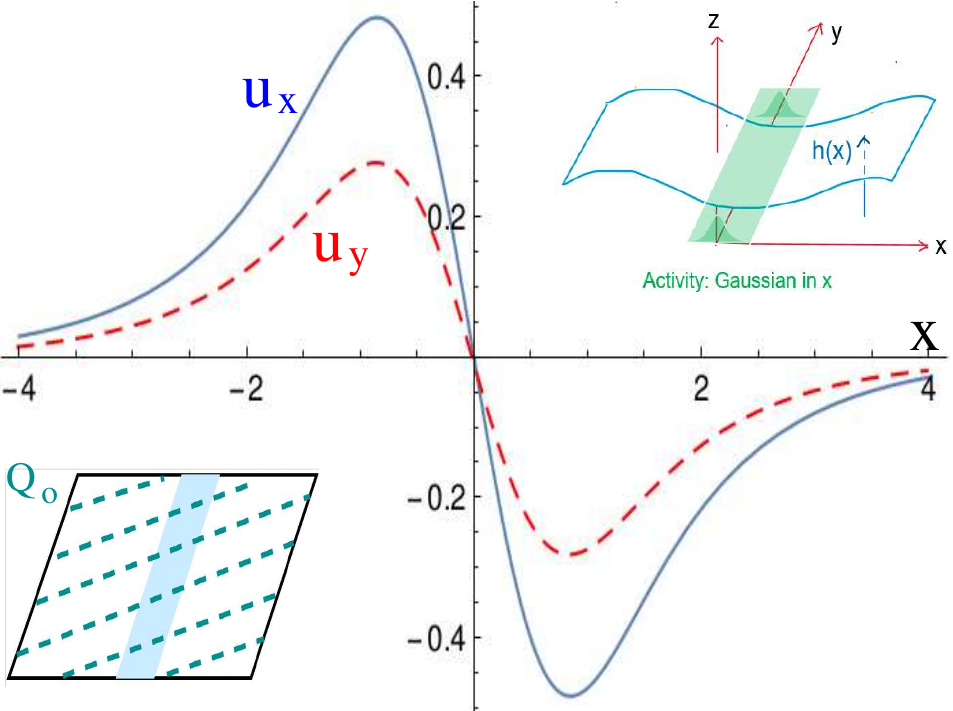} 
\caption{Theoretical $u_x(x)$ and $u_y(x)$,
obtained from Eq.\ref{eq.uxuy_q} via inverse Fourier transform. Inset on upper-right: geometry of membrane undulation. Inset at lower-left: nematic orientation in the reference state $Q_0$.  Parameters used: $2\theta_0=30^o, \alpha=\eta=1$, in arbitrary units.
}
\label{fig.velx-y}
\end{figure}

\begin{figure}
        \centering
    \includegraphics[width=8.5cm, height=6cm]{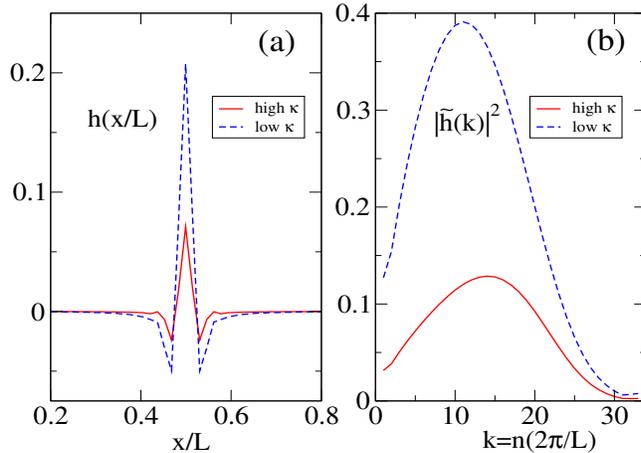} 
    \caption{Deformation of the interface (solutions of Eq.\ref{eq.h})  subjected to the compressive flow (Eq.\ref{eq.uxuy_q}) are shown. (a) and (b) show the height profile $h(x)$ and its power spectrum $|h(k)|^2$, respectively. Two different values of bending rigidity $\kappa$ were used; higher the $\kappa$ weaker the deformation. Initial condition for $h(x)$ was chosen to be random, yet the system reached steady state profile shown in (a). }
    \label{fig.buckle}
    \end{figure}

While our simulation reveals that a contractile ring is
essential for a sharp invagination, the relative importance of interface curvature and the compressive
flow are not clear. 
We therefore consider a simplified geometry, namely a  contractile, open surface subjected to a compressive flow (see Fig.\ref{fig.velx-y}-inset). 
The surface height profile is represented in Monge-Gauge as
$h(x,y)$. It models the interface between two phases
with $\phi=1$ and $\phi=-1$. The phase field relates to the
interface height $h(x,y)$ via $\phi=z-h(x,y)$ such that $\phi=0$ surface marks the interface. We work in the limit 
$|\vec \nabla h|\ll 1$. The interface is active and has a nematic field $Q_{ij}(\vec r)$ on it
(see inset of Fig.\ref{fig.buckle}). We consider a static inhomogeneous activity field $\zeta\Delta \mu (\vec r)$ which
is a function of $x$ only, given by $\zeta(\vec r)=\zeta_0 \exp(- x^2/2\beta^2)$, a gaussian profile peaked along $y$-axis (Fig.\ref{fig.buckle}-inset). The goal here is to explore whether this imposed
activity gradient, which creates a compressive stress $\sigma_{ij}$ toward the $y$ axis, and therefore induces a flow towards the middle of the surface, can generate surface instability. In order to focus on the convective
effects, we first ignore elastic stresses due to the gradients in nematic orientations, assume an 
interface friction ($-\alpha {\bf u}$) and consider the following reduced set of equations.
\begin{eqnarray}\label{eq.v}
\frac{\eta}{2} \nabla^2 u_i + \partial_j[\zeta Q_{ij}]= \alpha u_i \\
\frac{\partial \phi}{\partial t} + {\vec\nabla}.
(\phi \vec u) = -\delta F/\delta \phi\;.
\label{eq.phi}
\end{eqnarray}
Here $F=\frac{1}{2}\int [\kappa(\nabla^2 h)^2 +k_{\Phi}(\nabla h)^2]dA$ is a simplified Helfrich type surface free energy, where $\kappa$ and $k_{\Phi}$ are bending modulus and surface tension coefficient, respectively. We have ignored the dynamics of 
the $Q_{ij}$ field here and will explore possible convective instability of the flat surface with a fixed $Q_{ij}=Q^0_{ij}$ profile on it. Connecting phase to the height field, via $\phi=z-h(x,y)$  and using Lubrication approximation $|u_z|\ll |\vec u_{\perp}|$, where  $\vec u_{\perp}=(u_x,u_y$) \cite{howard-stone}, Eq.\ref{eq.phi} yields, 

\begin{equation}
\frac{\partial h}{\partial t} + \vec\nabla _{\perp}. (h\vec u_{\perp})
 = \kappa \nabla _{\perp}^4 h + k_{\Phi} \nabla _{\perp}^2 h \label{eq.h}
\end{equation}
Here $\vec\nabla _{\perp}\equiv (\partial_x,\partial_y)$  and variations along $z$ are ignored. We will first solve for the velocity field from Eq.\ref{eq.v} and then substitute the solution into Eq.\ref{eq.h} to investigate possible instability of the interface height. 

The 2nd term $\partial_j(\zeta Q_{ij})$ in Eq.\ref{eq.v}, although linear in $Q_{ij}$, has non-constant coefficient 
via the space dependence nature of the activity 
$\zeta (\vec r)$. To make progress we 
will consider an uniform $Q_{ij}$ field so that the spatial
derivative will act only on the  inhomogeneous activity field. For $i=x$
and $i=y$, the corresponding equations are,
\begin{equation}
\frac{\eta}{2} \partial_x^2 u_x +Q_{xx} \partial_x \zeta = \alpha u_x\;,\;\;\mbox{and}\;\;\;
\frac{\eta}{2} \partial_x^2 u_y +Q_{xy} \partial_x\zeta = \alpha u_y.
\label{eq.ux-y}
\end{equation}
Note, that a reference state where the nematic is oriented purely along x-axis and $Q_{xy}=\langle n_xn_y\rangle=0$, fails to produce any 
nonzero $u_y$. Our numerical simulation  
on the other hand, shows significant $u_{\theta}$ component near the 
equator (see Fig.\ref{fig.evolve}-c,d and Fig.\ref{fig.stress}-d), coupled with a nonzero nematic inclination $n_{\theta}$ 
in the two lobes (see Fig.\ref{fig.evolve}-a,b. 
For our simplified model this amounts to a nonzero $n_y$ component.
Motivated by this we consider a reference nematic field which 
maintains a finite angle $\theta_0$ with the x-axis i.e, 
$\hat n= (\cos\theta, \sin\theta)$. The resulting 
$Q^0_{ij}$ field reads
\[
Q^0 =\frac{1}{2}
\begin{pmatrix}
\cos 2\theta_0 & \sin 2\theta_0\\
\sin 2\theta_0 & -\cos 2\theta_0
\end{pmatrix}
\]

In Fourier space the solutions to Eq.\ref{eq.ux-y} are 
\begin{equation}
u_x (k_x)= \frac{ik_x Q^0_{xx}\zeta(k_x)}{\alpha + \eta k_x^2 },
\;
u_y (k_x)=\frac{ik_x Q^0_{xy}\zeta(k_x)}{\alpha + \eta k_x^2 }.
\label{eq.uxuy_q}
\end{equation}
The finite $Q^0_{xy}$ produces finite $u_y$ near y axis.
The corresponding analytical solutions are plotted in Fig.\ref{fig.velx-y} for $2\theta_0=30^o$, with $Q^0_{xx}=\sqrt 3/4$ and $Q^0_{xy}=-1/4$. This $u_y$ is qualitatively similar to motion in the azimuthal direction ($u_\theta$) observed near the furrow in our simulation, for 
random initial conditions. For in vivo cell divisions, 
this type  of actomyosin movement in the azimuthal direction, near the division furrow, has been reported \cite{naganathan-chiral,  pimpale2020cell}. 
    
Now turning to Eq.\ref{eq.h}, we note that the inhomogeneous velocity field (Eq.\ref{eq.uxuy_q}) in the convection term produces convolution in the Fourier space. Thus, it is difficult to perform linear stability analysis using a perturbation $\delta h$ over a flat interface. Instead, we solve for $h(x,t)$ numerically. When started from a random profile $h(x,0)$ with zero average, Eq.\ref{eq.h} leads to steady state surface profiles shown in Fig.\ref{fig.buckle}, for different values of the bending rigidity $\kappa$. Fig.\ref{fig.buckle}-a indicates onset of instability at the middle where the flow converges, however this small wave number instability (Fig.\ref{fig.buckle}-b) is arrested by the convective nonlinearity itself. At high wave numbers both the bending rigidity and surface tension suppress instabilities. 
This simplified model, however has $h\rightarrow -h$ symmetry, 
and is equally likely to buckle upward or downward. In contrast, the closed cortical interface, does not have 
up/down (or in/out) symmetry due to conservation of the cell volume and always invaginates inward. \\

In summary, the active shell model presented here, captured some important features of cytokinesis, namely, a) the spontaneous nematic ring formation, and b) the azimuthal components of the cortical flows. Our reduced analytical models pointed at nematic inclination ($Q_0$)
as possible driver of chiral motion, and the compressive equatorial flow as the driver of surface invagination. However, to keep the system of equations tractable, we could not include dynamics of myosin (stress regulators) and
instead assumed a time independent myosin concentration profile as a phenomenological input. The Dresden group has shown that such concentration  profile may emerge from instability of underlying reaction diffusion models, with mechano-chemical feedback, both in 
1D \cite{bois}, as well as on 3D closed, deforming, active vesicles \cite{l2-instability}. However, nematic nature of the stress regulators and azimuthally asymmetric velocity fields were not considered. 
Further, our model here does not address the late stage of cytokinesis within a confinement, e.g., within a hard shell (as in C elegans) \cite{sain-julicher, mainak-PRL} or within a confluent tissue, where the dividing lobes are stitched by membrane linkers (cadherin) to form a separating interface.


We thank Julia Yeomans for many helpful discussions and Frank Julicher for useful comments. We also thank Industrial Research and Consultancy Center (IRCC), IIT Bombay for financial support through SCPP grant.
\bibliographystyle{apsrev4-2}
\bibliography{ref}
\end{document}